\documentclass[aps,prl,twocolumn,groupedaddress]{revtex4-2}
\usepackage{graphicx,color}
\usepackage{amssymb}   
\usepackage{amsmath}
\DeclareMathOperator{\sign}{sgn}

\usepackage{epstopdf}
\usepackage{natbib}
\usepackage[colorlinks,citecolor=green,urlcolor=blue,bookmarks=false,hypertexnames=true]{hyperref} 
\usepackage{bm}
\usepackage{color}
\usepackage{braket}

\begin{document}
\title{Cu-substituted lead phosphate apatite as an inversion-asymmetric Weyl semimetal}
\author{Benjamin T. Zhou} \thanks{Corresponding author: benjamin.zhou@ubc.ca}
\author{Marcel Franz} \thanks{Corresponding author: franz@phas.ubc.ca}

\affiliation{Department of Physics and Astronomy \& Stewart Blusson Quantum Matter Institute,
University of British Columbia, Vancouver BC, Canada V6T 1Z4}

\begin{abstract}
Based on symmetry arguments and the latest density functional results for the copper-substituted lead phosphate apatite (`LK-99'), we show that, at the non-interacting level, the material is an inversion-asymmetric Weyl semimetal. A pair of Weyl nodes with opposite chiralities emerge at different energies in the vicinity of the time-reversal-invariant $\Gamma$ and ${\rm A}$ points of the 3D Brillouin zone. These are characterized by unusual Weyl charges of $C_{\rm W} = \pm 2$ and are connected by two branches of topologically protected Fermi arc states on surfaces parallel to the principal $c$-axis. We further study important effects of the atomic spin-orbit coupling on the band structure and the electronic properties of the material in general. Possible implications of the proposed band topology on the strong correlation physics are also discussed.    
\end{abstract}
\pacs{}

\maketitle

\emph{Introduction.}--- Achieving room-temperature superconductivity at ambient pressure is one of the ultimate goals of modern condensed matter research. In recent experiments on copper-substituted lead phosphate apatite Pb$_9$Cu(PO$_4$)$_6$O ~\cite{Lee-Kwon, Lee-Auh}, known also as `LK-99', tentative signatures of strong diamagnetism and low-resistance states have been observed. These results  hint at the possibility of room-temperature superconductivity and have ignited world-wide interest in  this family of  materials~\cite{Wu, Liu, Hou, Guo, Zhu, Timokhin, Kumar, Baskaran, Ya-Hui, Abramian}.

\begin{figure}
\centering
\includegraphics[width=0.5\textwidth]{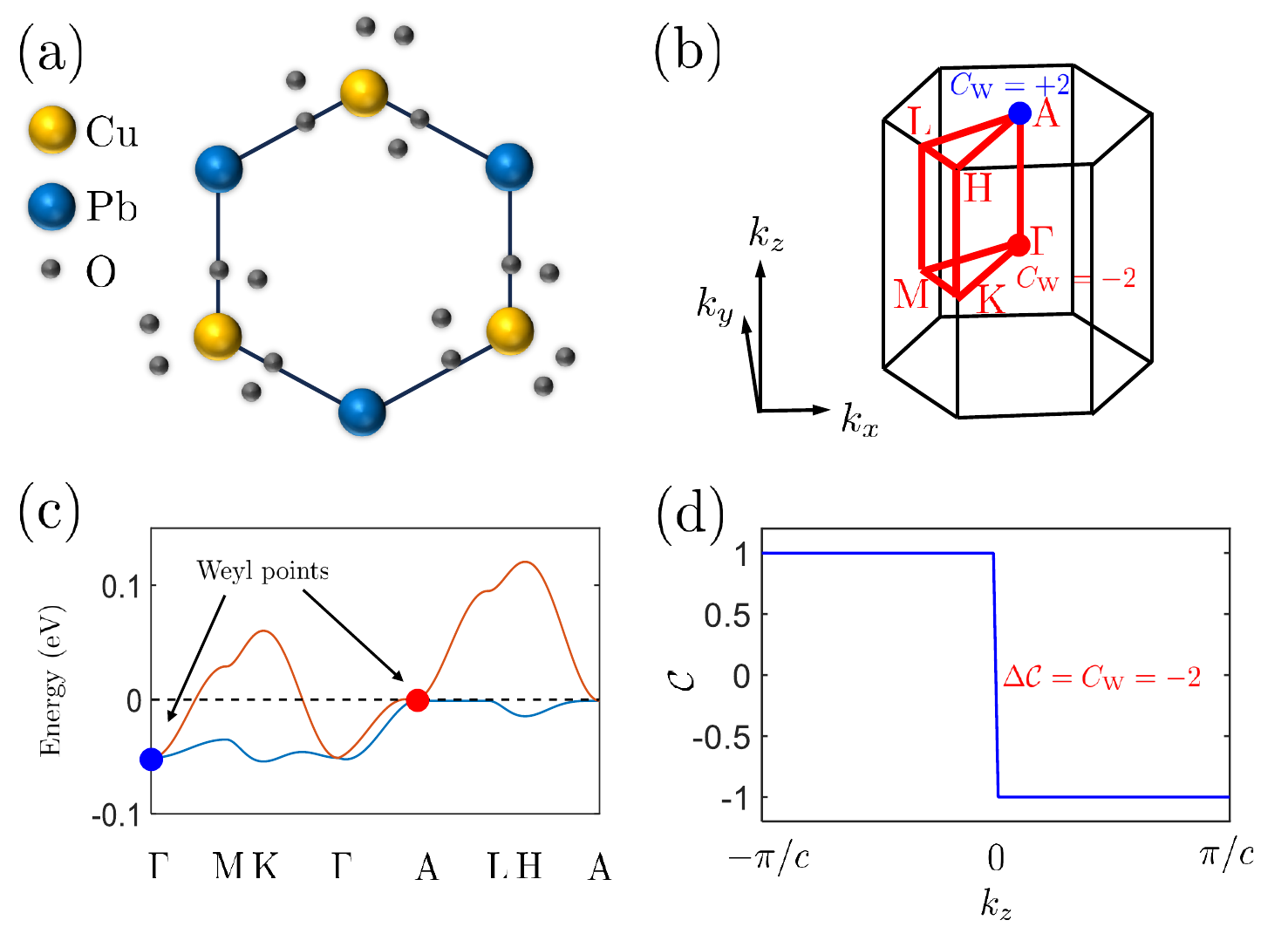}
\caption{(a) Schematic of the six-fold coordinated copper (Cu) and lead (Pb) atoms arranged in a triangular lattice. Mirror reflection is broken by the rotations of the triangles formed by oxygen (O) atoms.  (b) Hexagonal prismatic 3D Brillouin zone. (c) Energy bands obtained from two-band tight-biding model Eq.\ \eqref{eq:TB} with parameters obtained by fitting DFT bands in Ref.~\cite{Griffin}. (d) Chern number $\mathcal{C}$ as a function of  $k_z$.}
\label{FIG1}
\end{figure}

To illuminate physical mechanisms behind the observed phenomena, a microscopic theoretical understanding of the electronic properties of lead apatite is of critical importance. Following this line of thought several first-principle density functional theory (DFT) studies have been performed recently on LK-99~\cite{Griffin, LiangSi, Munoz, Lai}, which suggest that the stable crystal has a distorted trigonal prismatic structure with six-fold coordinated Cu atoms (Fig.\ \ref{FIG1}a). The relevant electronic states consist of two isolated bands stemming from the $d_{xz},d_{yz}$ orbitals of Cu with an overall bandwidth of $\sim 0.1$ eV and the Fermi level lying in the middle of the bands suggesting metallic behavior. The two isolated bands further exhibit interesting band crossing features at the time-reversal-invariant $\Gamma = (0,0,0)$ and ${\rm A} = (0,0,\pm \pi/c)$ points of the three-dimensional hexagonal prismatic Brillouin zone (BZ), while bandgaps are found in the rest of the BZ (Fig.\ \ref{FIG1}b).

With the electronic band structure and  the relevant atomic orbital compositions at hand, an important next step is to understand the symmetry and topological properties of the bands as well as the role of electron-electron interactions which are expected to be strong compared to the bandwidth.  In this Letter, we uncover an important topological aspect of the LK-99 band structure  --  we show that the band crossings near $\Gamma$ and A are Weyl points with opposite Weyl charges. Using a combined approach of symmetry analysis and microscopic modeling, we show that the emergence of Weyl points results from a combination of quadratic band touching enforced by the three-fold rotation symmetry ($\mathcal{C}_{3z}$) on the $\ket{d_{xz}} \pm i \ket{d_{yz}}$ doublet, and a finite band splitting along the $\Gamma-{\rm A}$ line caused by broken mirror symmetries. Together, these endow the crossing points with an unusual Weyl charge of $C_{\rm W} = \pm 2$ (Fig.\ \ref{FIG1}c-d). 

In accord with the bulk-boundary correspondence the two Weyl nodes with opposite charges are connected by a pair of topologically protected surface Fermi arcs which we visualize by a direct calculation.
%
We further study the important role of atomic spin-orbit coupling (SOC) on the  band topology and other electronic properties of the system, as well as discuss possible implications of the Weyl physics on the role of strong correlations in this narrow-band system. 

\emph{Effective spinless Weyl Hamiltonian.}--- For simplicity, we first consider a spinless model to capture the spin-polarized bands obtained by DFT. As shown in Ref.~\cite{Griffin}, the stable crystalline structure of LK-99 exhibits an approximate $C_{3v}$ point group symmetry. Given the relevant $d_{xz}, d_{yz}$-orbitals forming the two isolated bands, the doublet $\ket{d, \pm 1} = \ket{d_{xz}} \pm i \ket{d_{yz}}$ associated with orbital angular momenta $m_z = \pm 1$ forms the two-dimensional irreducible representation $E$ of $C_{3v}$. In addition, rotations in the triangles formed by oxygen (O) atoms break the vertical mirror plane $\sigma_v$ and reduce the point group from $C_{3v}$ to the chiral $C_{3}$ point group, which contains only the three-fold rotation $\mathcal{C}_{3z} \equiv e^{-i \frac{2\pi}{3} \tau_z}$ under the basis of $\ket{d, \pm 1}$ but no improper rotations. The $C_{3z}$-symmetry together with the spinless time-reversal $\mathcal{T}'=\tau_x\mathcal{K}$ dictates that up to lowest-order terms the spinless effective ${\bm k}\cdot{\bm p}$ Hamiltonians near time-reversal-invariant $\Gamma=\bm{Q}_{+}= (0,0,0)$, ${\rm A}=\bm{Q}_{-} = (0,0, \pm \pi/c)$ points can be written as
\begin{eqnarray}\label{eq:Weyl}
H_{W}^{\pm}(\bm{p}) = v_{\pm} [(p_x^2 - p_y^2) \tau_x - 2p_x p_y \tau_y] \pm v_z p_z\tau_z,
\end{eqnarray}
where momentum $\bm{p} = \bm{k}  - \bm{Q}_{\pm}$ is measured from $\bm{Q}_{\pm}$ points and $v_{\pm}$ denotes the velocity resulting from electron hopping within the $ab$-plane with $v_{+} \neq v_{-}$ in general. The $v_z p_z$ term originates from the breaking of mirror symmetry $\sigma_v$ due to rotations in O-triangles and causes the band splitting along the $\sigma_v$-invariant $\Gamma$-A line (Fig.\ \ref{FIG1}b), the $\tau_{\alpha=x,y,z}$ operate on the orbital subspace formed by $\ket{d, \pm 1}$. 

It is worth noting that the leading inter-orbital term is quadratic in the in-plane momentum $\bm{p}_{||} = (p_x, p_y)$ and has the same form as the well-known quadratic band touching in bilayer graphene~\cite{McCann1, McCann2}. The latter is known to generate a $2\pi$ Berry phase around the origin and results in a nonzero Chern number $\mathcal{C} = \sign(\Delta)$ in the quadratic band when a mass term $\Delta$ is introduced to gap out the band touching~\cite{FanZhang}. In Eq.\ \eqref{eq:Weyl}, the mass term $\Delta$ is given by the $v_z p_z$ term which changes sign as $p_z$ goes across the band touching points. This implies that the Chern number must change from $\pm 1$ to $\mp 1$ across the band touching at $\bm{Q}_{\pm}$, which indicates the presence of monopoles of charge $C_{\rm W} = \pm 2$~\cite{Wan, Ashvin, Binghai}. To confirm this result, we construct a symmetry-based realistic two-band tight-binding model in the Bloch basis $\ket{\bm{k}, d, \pm 1}$:
\begin{eqnarray}\label{eq:TB}
H_{\rm TB}(\bm{k})  =
\begin{pmatrix}
h_{++}(\bm{k}) &  h_{+-}(\bm{k})\\
h_{-+}(\bm{k}) & h_{--}(\bm{k})
\end{pmatrix},    
\end{eqnarray}
where $m, m' = \pm 1$ denotes the orbital index for $\ket{d, \pm 1}$ and details of the matrix elements $h_{mm'}(\bm{k})$ are presented in the Supplemental Material (SM)~\cite{SM}. The energy bands obtained from $H_{\rm TB}(\bm{k})$  are shown in Fig.\ \ref{FIG1}b, and are in excellent agreement with DFT bands. They also capture the effective quadratic Weyl Hamiltonian in Eq.\ \eqref{eq:Weyl} near $\bm{Q}_{\pm}$. The Chern number $\mathcal{C}$ as a function of $k_z$ is calculated using the eigenstates of $H_{\rm TB}(\bm{k})$ as presented in Fig.\ \ref{FIG1}d, which clearly shows the sudden change in $\mathcal{C}$ at $k_z = 0$ $(k_z =\pm\pi/c)$ and signifies Weyl charges of $C_{\rm W} = \pm 2$.

\begin{figure}
\centering
\includegraphics[width=0.5\textwidth]{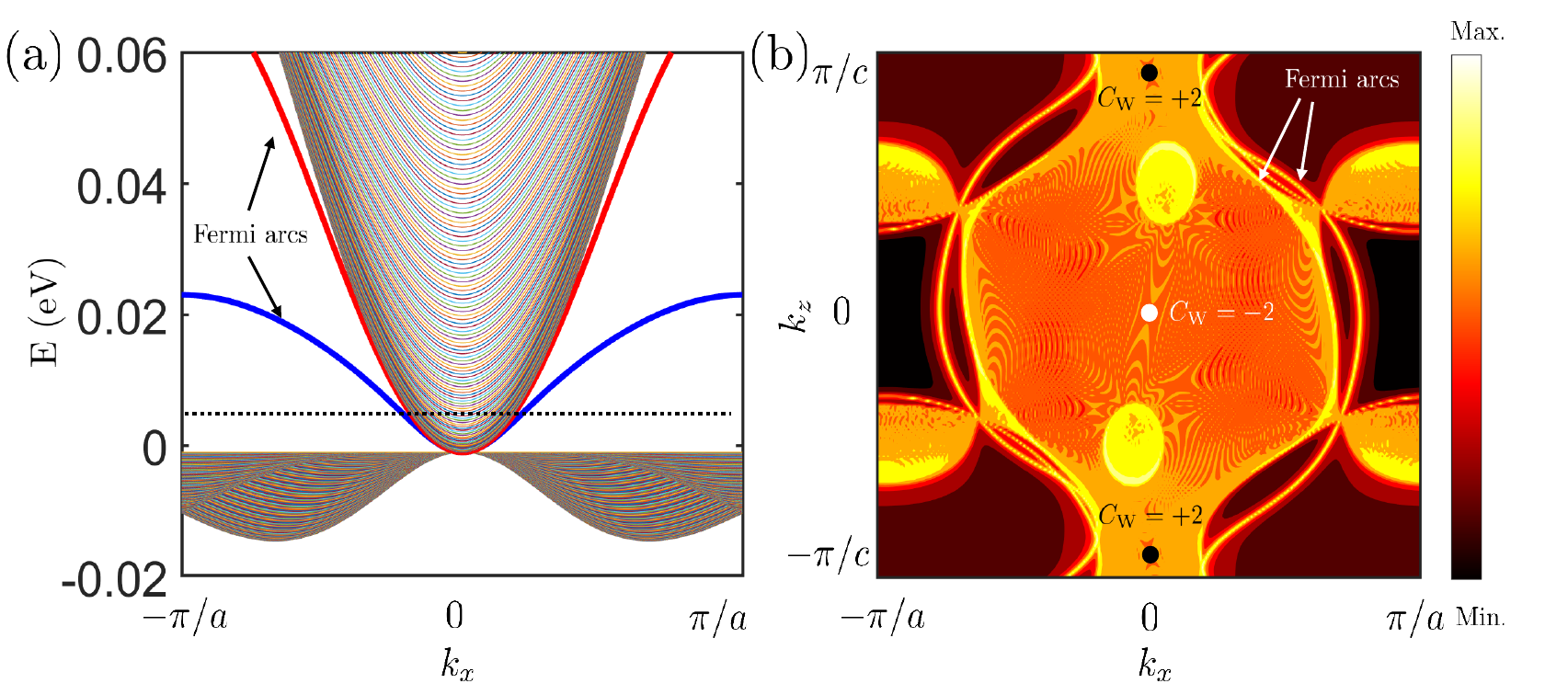}
\caption{(a) Energy spectrum at $k_z = \pi/c$ as a function of $k_x$ of an infinite slab with number of sites $N_y = 200$ along the $y$-direction. (b) Local density of states on the surfaces at $y = 0$ and $y = N_y$ in the $k_x - k_z$ plane obtained at energy $E = 5$ meV (dashed line in (a)). Color bar indicates the magnitude of local density of states on logarithmic scale.}
\label{FIG2}
\end{figure}

\emph{Topologically protected Fermi arcs.}-- As mandated by the bulk-boundary correspondence principle, Weyl points of opposite charges must be  connected by topologically protected gapless states, known as Fermi arcs~\cite{Wan, Ashvin, Binghai}, which live on surfaces where the projected Weyl charges do not cancel. As the Weyl nodes of opposite charges in LK-99 are located along the $\Gamma - {\rm A}$ line, Fermi arcs are expected to emerge on the side surfaces parallel to the $c$-axis. We demonstrate this explicitly by solving the tight-binding model in Eq.\ \eqref{eq:TB} in a slab geometry, infinite in the $xz$-plane with open boundaries terminated at $y=0$ and $y = L_y$ along the $y$-direction (see SM~\cite{SM} for details). 
The resulting energy spectrum at $k_z = \pi/c$, presented in Fig.\ \ref{FIG2}a, clearly shows two branches of Fermi arc states associated with Weyl charge of $C_{\rm W} = \pm 2$, separated from the bulk continuum and emanating from the projected Weyl point at $(k_x, k_z) = (0, \pi/c)$. 

To demonstrate that these states are predominantly localized on the surfaces, we further calculate the local density of states on the surfaces at $y = 0, L_y$ at energy $E = 5$ meV (indicated by dashed line in Fig.\ \ref{FIG2}a) throughout the entire surface Brillouin zone defined by conserved momenta $k_x$ and $k_z$ (Fig.\ \ref{FIG2}b). It is evident that the states connecting the two Weyl points are predominantly localized on the surface (brightness in color scale indicates the density of states). Note that due to the energy difference $\sim 50$ meV between Weyl points at $\rm \Gamma$ and $\rm A$ (Fig.\ \ref{FIG1}b), the bulk Fermi surface around $\Gamma$  is larger than that around  $\rm A$ when the Fermi level is close to $\rm A$ (Fig.\ \ref{FIG2}a). Hence the Weyl point $\bm{Q}_{+}$ is embedded in the bulk bands as shown Fig.\ \ref{FIG2}b.

\emph{Effects of atomic spin-orbit coupling.}--- We note that the emergence of Weyl points at the time-reversal-invariant $\rm \Gamma$ and $\rm A$ points in the spinless models (Eq.\ \ref{eq:Weyl}-\ref{eq:TB}) above is a result of the combination of spinless time-reversal $\mathcal{T}'$ and three-fold rotation $\mathcal{C}_{3z}$. However, the physical time reversal $\mathcal{T} = i\sigma_y \tau_x \mathcal{K}$ ($\sigma_{\alpha=x,y,z}$: spin Pauli matrices) involves the spin degrees of freedom and the true Kramers doublets are formed by two distinct pairs: $\{ \ket{d, +1, \uparrow}, \ket{d, -1, \downarrow}\}$ and $\{ \ket{d, +1, \downarrow}, \ket{d, -1, \uparrow}\}$, which indicates that the two-fold degeneracy at $\rm \Gamma$ and $\rm A$ within the same spin sector can generally be lifted when  $\mathcal{T}'$ is broken while $\mathcal{T}$ is respected. This happens if we include atomic spin-orbit coupling, which indeed seems relevant for LK-99 as evidenced by the energy difference of a few meVs found between spin-up and spin-down DFT bands~\cite{Griffin}. This suggests that the Weyl points, while being topological objects and immune to weak perturbations, are not necessarily pinned at the time-reversal-invariant points by any symmetry and can therefore be shifted by an appropriate perturbation. 

\begin{figure}
\centering
\includegraphics[width=0.5\textwidth]{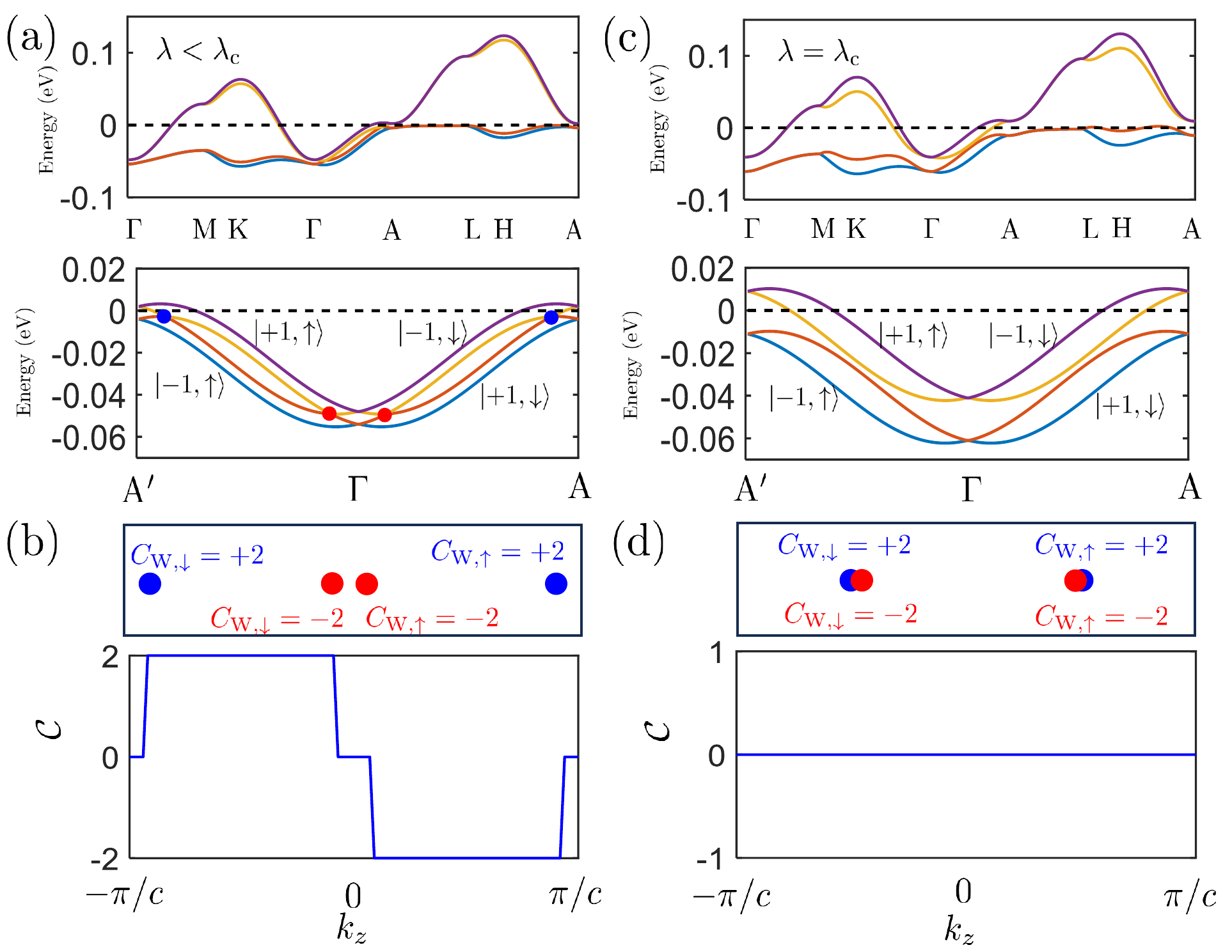}
\caption{(a) Energy bands with atomic SOC strength $\lambda = 6$ meV in the 3D BZ (upper panel) and along the $\rm A'$ - $\rm \Gamma$ - $\rm A$ line (lower panel) with $\rm A' = (0, 0 ,-\pi/c)$. Weyl points from spin-up and spin-down sectors are shifted. (b) Upper panel: schematic of Weyl point locations under finite SOC. Lower panel: Chern number as a function of $k_z$ with the same $\lambda$ in (a). (c) Energy bands with critical atomic SOC strength $\lambda_{\rm c} = 20$ meV in the 3D BZ (upper panel) and along the $\rm A'$ - $\rm \Gamma$ - $\rm A$ line (lower panel). Weyl crossings within the same spin sectors disappear. The remaining band crossing points between different spin sectors at $\rm \Gamma$, $\rm A$ are due to Kramers degeneracy, and are topologically trivial. (d) Upper panel: schematic of annihilation of Weyl points with opposite charges for each spin sectors at $\lambda_{\rm c} = 20$ meV. Lower panel: Chern number as a function of $k_z$ becomes uniformly zero at $\lambda_{\rm c}$.}
\label{FIG3}
\end{figure}

It can be shown (see SM~\cite{SM} for details) that in the relevant subspace spanned by  $\{ \ket{d, m=\pm1, \sigma = \uparrow, \downarrow}\}$, the atomic SOC takes the simple form of 
\begin{eqnarray} \label{eq:SOC}
H_{\rm SOC} = \frac{\lambda}{2} \tau_z \otimes \sigma_z,    
\end{eqnarray}
where $\lambda$ characterizes the SOC strength.   Fig.\ \ref{FIG3}a shows that SOC causes an energy level splitting  of $\Delta E = \lambda$ at $\rm \Gamma$ and $\rm A$ between $\ket{d, m=+1, \sigma}$ and $\ket{d, m=-1, \sigma}$ for a given spin $\sigma$. Note that because $H_{\rm SOC}$ in Eq.\ \eqref{eq:SOC} is diagonal in the spin basis the two spin sectors remain decoupled in the presence of SOC. However, due to the splitting induced by $\lambda \neq 0$, the Weyl points from each spin sector are shifted away from their high-symmetry positions to generic points $k_{z,0} \neq 0, \pi/c$ along the $\rm \Gamma-A$ line as shown in Fig.\ \ref{FIG3}a,b. 

Notably, for large enough $\lambda$ the Weyl points with opposite charges can collide and annihilate, rendering the system topologically trivial (Fig.\ \ref{FIG3}d). In the present model this occurs for $\lambda_{\rm c} \simeq 20$ meV.  Given the spin splitting of $<10$ meV found in the DFT calculations~\cite{Griffin}, our results suggest that LK-99 is within the $\lambda < \lambda_{\rm c}$ regime where the nontrivial Weyl physics remains valid in the presence of atomic SOC.

On the other hand, we note that even in the strong SOC limit with $\lambda > \lambda_{\rm c}$ where the system is topologically trivial, the strong SOC effect shown in Fig.\ \ref{FIG3}b has important consequences on the electronic properties. In fact, the form of the SOC splitting shown in Fig.\ \ref{FIG3}b is reminiscent of the well-known spin-valley locking~\cite{Xiao, ZYZhu, Kormanyos} or Ising SOC~\cite{Benjamin, Curtis} in atomic layers of transition-metal dichalcogenides (TMDs). In particular, in the high hole (electron) doping limit where only the $K$-valleys ($H$-valleys) are accessed, the strong spin splitting induced by SOC leads to similar coupled spin-valley physics in TMDs which can give rise to long spin and valley life times as the flip of spin and valley indices must occur simultaneously. These results suggest that LK-99 under strong SOC limit may have potential applications in spintronics and valleytronics \cite{Xu, Zefei}.

\emph{Implications for diamagnetism, low-resistivity and correlation physics.}--- Having established the Weyl physics in the above sections, we now discuss its possible implications for the recent experimental observations interpreted as signatures of high-$T_{\rm c}$ superconductivity in LK-99. First, we note that strong diamagnetism need not originate exclusively from the Meisser effect in a superconductor -- a well-known alternative is atomically thin graphene, where the orbital diamagnetic susceptibility in principle diverges when the Fermi level lies at the charge neutrality point with either linear or quadratic band touching~\cite{McClure, Koshino1, Koshino2, Ando, Vallejo}. The origin of this effect can be traced back to the formation of zero-energy Landau levels in monolayer and bilayer graphene under applied magnetic field~\cite{McCann1, Yuanbo, Geim}: electrons in occupied states need to increase their energy to join the zero-energy Landau level, which then increases the total energy of the system. Therefore, it is energetically favorable for electrons to partially expel the external field via their orbital motion. Similar physics is also known to occur in systems with three-dimensional Dirac spectrum where zero-energy Landau levels can also emerge~\cite{Marcel, Suetsugu, XiangYuan}.

As we discussed above, bulk Weyl fermions described by Eq.\ \eqref{eq:Weyl} can be regarded as multiple copies of bilayer graphene where zero-energy Landau level is expected to emerge under applied magnetic field. This suggests that a similar mechanism of Landau diamagnetism could be relevant to the microscopic origin of the strong diamagnetic effects found in some experiments on LK-99~\cite{Lee-Auh, Wu}.

Moreover, we note that low-dissipation electronic transport can also occur via topologically protected boundary modes due to suppressed back-scattering from spatial separation of left- and right-moving carriers~\cite{Buttiker, Klitzing, Cuizu}. In the context of Weyl semimetals, the currents carried by Fermi arcs, while not completely immune to dissipation due to scattering processes involving bulk channels~\cite{Gorbar}, have been shown to be remarkably disorder-tolerant and often lead to ultra-high electron mobility when the surface transport via Fermi arcs dominates \cite{Resta, ChengZhang}. In fact, as shown in Fig.\ \ref{FIG2}b, Fermi arcs can cover a large part of the surface Brillouin zone. This indicates a large number of Fermi arc channels, which could potentially explain the low-dissipation transport reported in LK-99 by some groups~\cite{Hou}.

To fully address the issues raised above, detailed calculations of the orbital diamagnetic effects and mesoscopic transport in LK-99  must be performed. While this lies beyond the scope of the current work we note that  a lattice model is a necessary starting point for the study of these effects. Our tight-binding model Eq.\ \eqref{eq:TB} captures the key features of the LK-99 band topology and could thus enable these calculations as well as serve as a basis for the study of strong correlation effects.

On the other hand, while our proposed Weyl physics suggests alternative explanations for the reported experimental signatures of superconductivity, it is important to note that the nontrivial band topology established in this work does not rule out the possibility of high-$T_{\rm c}$ superconductivity in this family of materials. On the contrary, Weyl physics opens even richer possibilities for the superconducting states as the interplay between correlations and topology often leads to exotic phases of matter. In particular, distinct from conventional superconductors with trivial normal-state band topology, the existence of protected surface states opens a new possibility of superconducting instability nucleated at the surface. 
In fact, recent studies have suggested the possibility that the $T_{\rm c}$ of surface superconductivity in a Weyl semimetal can actually exceed that in the bulk~\cite{Baenitz, CeHuang, Schimmel, Nomani}. If the high $T_{\rm c}$ superconductivity were eventually confirmed in LK-99, our results would be compatible with possible high-$T_{\rm c}$ superconductivity driven by the Fermi arc states.

Should the possible electron pairing occur alternatively in the bulk, the nontrivial band topology, particularly the Berry curvature generated by the Weyl points, also impose important topological constraints on the pairing symmetry which usually implies unconventional (non-$s$-wave) pairing~\cite{Alidoust, YiLi}. Even in the strong SOC limit with $\lambda > \lambda_{\rm c}$ where the bulk topology becomes trivial, the strong SOC splitting has consequences for the superconducting states as the breaking of spin SU(2) symmetry in the non-centrosymmetric superconductor would necessarily imply mixing between spin-singlet and spin-triplet pairing~\cite{Rashba, Sigrist}, which often leads to topological superconductivity~\cite{Wenyu, Hsu} and has potential applications for superconducting spintronics~\cite{Linder}. 

\emph{Conclusions.}-- Much remains unknown about the LK-99 family of materials studied in recent experiments. Perhaps most importantly it is not yet clear  whether Pb$_9$Cu(PO$_4$)$_6$O is the relevant stable crystal structure present in experimental samples. Indeed, a recent DFT study \cite{Jiang} suggested Pb$_9$Cu(PO$_4$)$_6$(OH)$_2$ to be a more likely candidate. Nevertheless, the latter compound shows very similar electronic structure near the Fermi level and our analysis and conclusions remain applicable with minor modifications. 

On the theory side some DFT results suggest the narrow bands to be spin-polarized \cite{Griffin,Jiang}, making the material an unlikely candidate for a high-temperature superconductor. Estimates of the copper on-site repulsion parameter $U$ further indicate that LK-99 could be in a very strongly correlated limit with $U\gg w$, the bandwidth \cite{LiangSi}. It is therefore unclear how electrons in narrow bands  avoid forming a large-gap Mott insulator at integer filling suggested by the chemical formula. Clearly, more experimental work is needed to gain insight into these issues. On the other hand, our analysis of the SOC effects on band topology remain applicable to spin-polarized bands: since the atomic SOC in Eq.\ \eqref{eq:SOC} is block-diagonal in spinor basis, the SOC effect in each set of spin-polarized bands remains the same as those shown in Fig.\ \ref{FIG3}. 

We close by noting that results presented in this work are relevant whether or not the material studied in Refs.\ \cite{Lee-Kwon, Lee-Auh} ends up being a supercondutor, or even contains LK-99 as a major ingredient. They establish an interesting band topology in Pb$_9$Cu(PO$_4$)$_6$O with two second-order Weyl points located very close to the Fermi level. By contrast most other Weyl semimetals known today exhibit a large number of Weyl points that tend to be buried among  trivial bands far away from the Fermi level, making experimental observation of fundamental phenomena (e.g.\ Fermi arcs, chiral anomaly) difficult. In this sense Pb$_9$Cu(PO$_4$)$_6$O/(OH)$_2$ may furnish a convenient realization of the minimal model Weyl semimetal which could facilitate experimental studies simply not feasible with existing materials.  

{\em Note. --} During the preparation of this manuscript we became aware of recent preprints (Ref.~\cite{Scaffidi, Hirschmann}) which reported minimal tight-binding models of LK-99 and Ref.~\cite{Hirschmann} also mentioned the existence of Weyl points in the bulk energy band structure.

{\em Acknowledgement. --} The authors are indebted to J. Berlinsky, D. A. Bonn, A. Damascelli, C. Felser, A. Hallas, P. Kim and V. Pathak for insightful discussions and correspondence. The work presented here 
was supported by NSERC, CIFAR and the Canada First Research Excellence Fund, Quantum Materials and Future Technologies Program.

\clearpage

\onecolumngrid
\begin{center}
\textbf{\large Supplemental Material for\\ ``Cu-substituted lead phosphate apatite as an inversion-asymmetric Weyl semimetal''} \\[.2cm]
Benjamin T. Zhou$^{1}$ and  Marcel Franz$^{1}$\\[.1cm]
{\itshape ${}^1$Department of Physics and Astronomy \& Stewart Blusson Quantum Matter Institute,
University of British Columbia, Vancouver BC, Canada V6T 1Z4}	\\
\end{center}
\setcounter{equation}{0}
\setcounter{section}{0}
\setcounter{figure}{0}
\setcounter{table}{0}
\setcounter{page}{1}
\makeatletter
\renewcommand{\theequation}{S\arabic{equation}}

\section*{I. Derivation of two-band tight-binding model based on group theory}

According to Ref.\ \cite{Griffin} the Cu-substituted apatite has six-coordinated Cu and Pb sites with distorted trigonal prismatic coordination and exhibits approximate $C_{3v}$ point group symmetry. Given the relevant $d_{xz}, d_{yz}$-orbitals for the two flat bands, we consider the chiral basis with magnetic quantum number $m_z = \pm 1$: $\ket{d, \pm 1} = \ket{d_{xz}} \pm i \ket{d_{yz}}$, which forms the two-dimensional irreducible representation $E$ of $C_{3v}$. In the basis of $\ket{d, \pm 1}$, the generators $C_{3z}$ and $M_{x}$ of $C_{3v}$ point group are represented by
\begin{eqnarray}
 D(C_{3z}) = 
 \begin{pmatrix}
 \omega_{-} & 0\\
 0 & \omega_{+}
 \end{pmatrix},
 && D(M_x) = 
  \begin{pmatrix}
 0 & -1\\
 -1 & 0
 \end{pmatrix},
\end{eqnarray}
where $\omega_{\pm} = e^{\pm i 2\pi/3}$, and the general form of the two-band tight-binding Hamiltonian $H(\bm{k})$ is given by
\begin{eqnarray}
H(\bm{k}) = 
\begin{pmatrix}
h_{+}(\bm{k}) &  h_{+-}(\bm{k})\\
h_{-+}(\bm{k}) & h_{-}(\bm{k})
\end{pmatrix}.
\end{eqnarray}
Given the symmetry constraints imposed by $C_{3v}$ 
\begin{eqnarray}
D(C_{3z})H(\bm{k})D^{\dagger}(C_{3z})&=&H(C_{3z}\bm{k}),\\\nonumber
D(M_{x})H(\bm{k})D^{\dagger}(M_{x})&=&H(M_x\bm{k}),
\end{eqnarray}
the matrix elements in $H(\bm{k})$ must satisfy:
\begin{eqnarray}
h_{\pm}(\bm{k})&=&h_{\pm}(C_{3z}\bm{k}),\\\nonumber
h_{+-}(\bm{k})&=&\omega^2_- h_{+-}(C_{3z}\bm{k}),\\\nonumber
h_{-+}(\bm{k})&=&\omega^2_+ h_{-+}(C_{3z}\bm{k}),\\\nonumber
h_{\pm}(\bm{k}) &=& h_{\mp}(M_x\bm{k}),\\\nonumber
h_{+-}(\bm{k}) &=& h_{-+}(M_x\bm{k}),
\end{eqnarray}
Consider Wannier orbitals of $d_{xz} \pm id_{yz}$ characters on the triangular lattice formed by Cu-atoms with nearest-neighbor hopping. The matrix elements in $H(\bm{k})$ have the following form
\begin{eqnarray}\label{eq:TBS}
h_{\pm}(\bm{k})&=& E_0 + 2t_0 C(\bm{k}_{||}) \pm 2 u_0 S(\bm{k}_{||}) \\\nonumber
               &+& 2 g_0 C_{z}(k_z) \pm 2 u_1 S(\bm{k}_{||})C_{z}(k_z)\\\nonumber
                &\pm& 2 g_{0z} S_z(k_z),\\\nonumber
h_{+-}(\bm{k})&=& 2t_1 C_{-}(\bm{k}_{||}) + 2 g_1 C_{-}(\bm{\bm{k}_{||}})C_{z}(k_z),\\\nonumber
h_{-+}(\bm{k})&=& 2t_1 C_{+}(\bm{k}_{||}) + 2 g_1 C_{+}(\bm{\bm{k}_{||}})C_{z}(k_z),
\end{eqnarray}
where $\bm{k}_{||} = (k_x, k_y)$ denotes the in-plane momentum, $t_0$ ($t_1$) is the in-plane intra-orbital (inter-orbital) nearest-neighbor hopping amplitude, $g_0$ is the normal intra-orbital hopping along the $c$-axis, $g_1$ is the inter-orbital inter-layer hopping. As we discussed in the main text, the rotations of triangles formed by oxygen atoms further breaks mirror symmetry $\sigma_{v} = M_{x}$, which reduces the point group to $C_3$. This broken symmetry allows the presence of $\pm 2 g_{0z} S_z(k_z)$ in $h_{\pm}(\bm{k})$, which reduces to the $v_z p_z \tau_z$ term in Eq.\ (1) of the main text. The basis functions in Eq.\ \eqref{eq:TBS} are 
\begin{eqnarray}
C(\bm{k}_{||})&=& \sum_{j=1,2,3} \cos(\bm{k}_{||}\cdot \bm{R}_{2j-1}),\\\nonumber
S(\bm{k}_{||})&=& \sum_{j=1,2,3} \sin(\bm{k}_{||}\cdot \bm{R}_{2j-1}),\\\nonumber
C_{\pm}(\bm{k}_{||})&=& \sum_{j=1,2,3} \omega^{j-1}_{\pm}\cos(\bm{k}_{||}\cdot \bm{R}_{2j-1}),\\\nonumber
C_z(k_z)&=& \cos(k_z c), S_z(k_z) = \sin(k_z c)
\end{eqnarray}
with $\bm{R}_{j=1,2,3} C^{j-1}_{3z} \bm{a}_1$ and $\bm{a}_1 = (a,0,0)$ the in-plane unit lattice vector. The symmetry properties of the basis functions are summarized as follows
\begin{eqnarray}
C(C_{3z}\bm{k}_{||})&=&  C(\bm{k}_{||}),\\\nonumber
S(C_{3z}\bm{k}_{||})&=&  S(\bm{k}_{||}),\\\nonumber
C_{\pm}(C_{3z}\bm{k}_{||})&=&  \omega_{\pm} C_{\pm}(\bm{k}_{||}),\\\nonumber
C(M_x\bm{k}_{||})&=&  C(\bm{k}_{||}),\\\nonumber
S(M_x\bm{k}_{||})&=&  -S(\bm{k}_{||}),\\\nonumber
C_{\pm}(M_x\bm{k}_{||})&=&  C_{\mp}(\bm{k}_{||}).
\end{eqnarray}

In the vicinity of the time-reversal-invariant momenta $\bm{\Gamma} = (0,0,0)$ and $\bm{A} = (0,0,\pi/c)$, the basis functions $C_{\pm}(\bm{k}_{\\}) \simeq -\frac{3a^2}{8}k^2_{\mp}$, where $k_{\pm} = k_x \pm i k_y$. Therefore, the effective ${\bm k}\cdot{\bm p}$ Hamiltonian near $\bm{Q}_{+}=\bm{\Gamma}$, $\bm{Q}_{-} = \bm{A}$ can be written as:
\begin{eqnarray}\label{eq:WeylS}
H_{W}(\bm{Q}_{\pm} + \bm{p}) = v_{\pm} [(p_x^2 - p_y^2) \tau_x - 2p_x p_y \tau_y] \pm v_z p_z\tau_z,
\end{eqnarray}
which coincides with  Eq.\ (1) of the main text, if we identiify $v_{\pm} = -\frac{3a^2 (t_1 \pm g_1)}{4}$, $v_z = 2g_{0z}c$. \\

\begin{table}[tp] 
\caption{Parameters in units of eV used in $H_{\rm TB}(\bm{k})$ (Eq. 2 of the main text) for LK-99 obtained by fitting to DFT bands in Ref.~\cite{Griffin}.}
\centering
\begin{tabular}{c c c c c c c c}
\hline \hline
 $E_0$ & $t_0$ &  $t_1$ & $u_0$ &  $g_0$  &  $u_1$ & $g_1$ & $g_{0z} $\tabularnewline
\hline 
 0.01 & -0.006  & -0.002 & -0.012 & -0.0125 & 0.001 & 0.01 & 0.004 \tabularnewline
\hline\hline
\end{tabular}
\label{table:01}
\end{table} 

\section*{II. Details of numerical calculation of Fermi arc states in an infinite slab geometry}

Here we present details of the numerical calculation of Fermi arc states and local spectral density in Fig.\ 2 of the main text. To account for the open boundaries along the $y$-direction, we perform a partial Fourier transform: $c^{\dagger}_{\bm{k}, m} = \frac{1}{\sqrt{N_y}} \sum^{N_y}_{j=1} e^{i k_y j b} c^{\dagger}_{j, k_x, k_z, m}$, where $b = \sqrt{3}a/2$. The Hamiltonian in terms of site index $j, j'$  along the y-direction is written as
\begin{eqnarray} 
 \mathcal{H} = \sum_{\braket{j,j'}} \sum_{k_x, k_z} \sum_{m, m'} c^{\dagger}_{j, k_x, k_z, m} h_{jj', mm'}(k_x, k_z)  c_{j', k_x, k_z, m'},
\end{eqnarray}
where the matrix elements $h_{jj', mm'}(k_x, k_z)$ are non-vanishing only for $j = j'$ and $|j-j'| = 1$. For $j=j'$, the matrix elements are given by
\begin{eqnarray} \label{eq:onsiteterms}
h_{jj,\pm}(k_x, k_z) &=&  E_0 + 2 g_0 C_z(k_z) \pm 2 g_{0z} S_z(k_z)\\\nonumber
                     &+&  2 t_0\cos(k_x a) \\\nonumber
                     &\pm& 2 [u_0 + u_1 C_{z}(k_z)] \sin(k_x a), \\\nonumber
h_{jj,+-}(k_x, k_z) &=&  2[t_1 + g_1 C_z(k_z)]\cos(k_x a), \\\nonumber
h_{jj,-+}(k_x, k_z) &=&  2[t_1 + g_1 C_z(k_z)]\cos(k_x a).
\end{eqnarray}

Terms with $j' = j-1$ are related to those with $j' = j+1$ through Hermitian conjugation; thus it suffices to work out the $j' = j+1$ case:
\begin{eqnarray} \label{eq:NNterms}
h_{j,j+1,\pm}(k_x, k_z) &=& 2 t_0 \cos(\frac{k_x a}{2})\\\nonumber
                     &\mp& 2 [u_0 + u_1 C_{z}(k_z)] \sin(\frac{k_x a}{2}), \\\nonumber
h_{j,j+1,+-}(k_x, k_z) &=&  2[t_1 + g_1 C_z(k_z)] \cos(\frac{k_x a}{2} + \frac{2\pi}{3}),  \\\nonumber
h_{j,j+1,-+}(k_x, k_z) &=&  2[t_1 + g_1 C_z(k_z)] \cos(\frac{k_x a}{2} - \frac{2\pi}{3}).  
\end{eqnarray}
By diagonalizing $\mathcal{H}$ we can obtain the spectrum shown in Fig.\ 2a of the main text. In Fig.\ 2b, we obtain the density of states on the surface via the spectral function
\begin{eqnarray}
A_{\rm surf}(k_x,k_y) = -\frac{1}{2\pi} {\rm{Im}}[G_{\rm surf}^{R}(k_x, k_z) -   G_{\rm surf}^{A}(k_x, k_z)],
\end{eqnarray}
where $G_{\rm surf}^{R}$ ($G_{\rm surf}^{A}$) are the surface retarded (advanced) Green's function of $\mathcal{H}$. In our calculation for Fig.\ 2 we set $N_y = 200$ and the surface Green's functions in $A_{\rm surf}(k_x,k_y)$ average over 5 atomic sites measured from the boundaries.

\section*{III. Atomic spin-orbit coupling for $d_{xz}, d_{yz}$-orbitals}

The general form of atomic spin-orbit coupling is written as
\begin{equation}
 H_{\rm SOC} = \frac{\lambda}{2} \bm{L} \cdot \bm{S},   
\end{equation}
where $\bm{L}$ and $\bm{S}$ denote the orbital and spin angular momentum operators, respectively. Given the $m_z = \pm 1$ orbitals relevant for the two isolated bands, we note that $\braket{m_z = \pm 1| L_{x,y}|m_z = \pm 1} = 0$. Thus the in-plane orbital angular momentum matrix vanishes for $d_{xz}, d_{yz}$ (up to leading order perturbations), which allows us to consider $L_z$ only. In the basis of $\ket{d, +1}, \ket{d, -1}$, we have 
\begin{equation}
L_z = 
\begin{pmatrix}
1 & 0\\
0 & -1
\end{pmatrix}
\equiv \tau_z, 
\end{equation}
where $\tau_z$ is the Pauli matrix acting on the two orbitals as defined in Eq.\ 1 of the main text. Thus, the final form of SOC is written as $H_{\rm SOC} = \frac{\lambda}{2} \tau_z \sigma_z$ as given by Eq.\ (3) of the main text.

\end{document}